\documentclass[%
 reprint,
superscriptaddress,
 amsmath,amssymb,
 aps,
]{revtex4-2}
\usepackage{xcolor}
\usepackage{graphicx}% Include figure files
\usepackage{dcolumn}% Align table columns on decimal point
\usepackage{bm}% bold math

\begin{document}

\title{Invariant-mass spectroscopy in projectile fragmentation reactions}

\def\WUCHEM{Department of Chemistry, Washington University, St. Louis, Missouri 63130, USA}
\def\WUPHYS{Department of Physics, Washington University, St. Louis, Missouri 63130, USA}
\

\author{R.J. Charity}
\affiliation{\WUCHEM}
\email{charity@wustl.edu}

\author{L.G. Sobotka}
\affiliation{\WUCHEM}
\affiliation{\WUPHYS}

\date{January 2023}
\date{\today}% It is always \today, today,
\begin{abstract}
The fragmentation of a projectile into a number of pieces can lead to the creation of many resonances in different nuclei. We discuss  application of the invariant-mass method to the products from such reactions to find some of the most exotic resonances located furthest beyond  the proton drip line. We show examples from fragmentation of a fast $^{13}$O beam including the  production of  the newly identified $^9$N resonance. In extracting resonance parameters from invariant-mass spectra, accurate estimates of the background from non-resonant prompt protons are needed.  This is especially important in determining the widths of wide resonances typically found at the edge of the chart of nuclides.  An event-mixing recipe, where the mixed events have reduced weighting for the smaller invariant-masses, is proposed to describe this background. The weighting is based on the measured correlations of heavier hydrogen isotopes with the resonances or the projectile residues. 
\end{abstract}

\maketitle

\section{Introduction}

Exploration of the nuclear landscape beyond the drip lines allows us to explore the most exotic nuclear states with the largest mismatch between the number of protons and neutrons. The invariant-mass method, where all the decay products are detected, is often used to find such exotic resonances and fits to the experimental invariant-mass distributions give their centroids and intrinsic widths. Many of the states beyond the drip lines decay by multi-nucleon emission and the invariant-mass method allows for the correlations between the nucleons to be determined \cite{Webb:2019a}. These correlations can give information on the structure of the parent state \cite{Wang:2021,Wang:2022}.

With fast radioactive beams, one and two-nucleon knockout reactions using light target nuclei (usually $^9$Be or $^{12}$C) are often used to create even more exotic species. 
While the arguments presented here have consequences for levels beyond both the proton and neutron drip lines, this work will concentrate on exploring just the proton-rich states.
One and two-neutron knockout reactions can be  considered as clean ways to search for proton resonances beyond the drip line. The first-step, direct knockout process produces no protons thus only delayed protons from second step  (resonant decay) are detected.

However if we restrict ourselves to such ``clean'' reactions, then we are limited to how far one can explore beyond the proton drip line. Also in the invariant-mass method, such clean processes cannot be completely separated from ``dirty'' processes.  For instance, the yield of  $p$+($Z_{projectile}$-1,$A_{projectile}$-2) events can have contributions from delayed protons produced in the  decay of a resonance following a 1$n$ knockout reaction
 or from events where both a $p$ and $n$ were promptly removed from the projectile in the  first-step. If one tries to use three-neutron knockout reactions to explore further beyond the drip line, do these ``dirty'' process overwhelm the the very small yields expected for direct 3$n$ removal? Perhaps it is more advantageous to make these more exotic states via such ``dirty'' reactions.  

The drip lines show strong odd-even staggering induced by nucleonic superfluidity.  In making these more exotic species, even-Z projectiles located just inside of the proton drip line thus have an advantage over their neighboring odd-Z cousins as they start out with larger proton-to-neutron imbalances.  Projectile fragmentation reactions where a proton and multiple neutrons are promptly liberated in the first step may offer a way to populate the the extreme $Z_{projectile}-1$ isotopes with more reasonable yields. One can also contemplate reactions where multiple protons and neutrons are removed from the projectile. However,
does the background from the prompt protons overwhelm the resonant peaks from the delayed protons in the invariant-mass spectrum? Even if  resonant peaks are visible, to reliably extract their intrinsic widths, the shape of the background from these prompt protons should be known. This is especially true for the very wide states which are expected as we progress further beyond the proton drip line.

In this work, we will explore these issues using data from an experiment with a $^{13}$O beam. Results from this experiment associated with $^{12}$O and $^{11}$O states produced from the ``clean'' one and two-neutron removal reactions has already been published \cite{Webb:2019,Webb:2019a,Webb:2020} and well as states in $^{10,11,12}$N,
$^{11}$C and $^{10}$C using more ``dirty'' projectile fragmentation processes \cite{Webb:2019a,Charity:2020,Charity:2021}. We will revisit some of the these cases with new insight and finally discuss the background in the very exotic case of $^{9}$N which decays by emitting five protons and is produced via the prompt removal of $p$+3$n$, in total, from the projectile. We will also present some data from a second experiment using a $^{9}$C beam with the same detector configuration, same target, and similar beam energy. 

\section{Experimental Method}
\label{sec:method}
The experimental data were obtained using  fast secondary beams produced at the Superconducting Cyclotron Laboratory at Michigan State University.  In all cases, the secondary beams were incident on the same 1--mm-thick 
$^9$Be target and the decay products were detected in the High Resolution Array (HiRA) 
\cite{Wallace:2007} with nearly identical configurations covering the same angular 
regions of the resonant-decay products ($\approx$2$^\circ$-12$^\circ$). In these experiments, 
HiRA consisted of 14 Si-CsI(Tl) $E$-$\Delta E$ telescopes. Each telescope contained a 
double-side Si strip detector which provided position (scattering angle) information 
and behind this was placed a 2$\times$2 array of  CsI(Tl) crystals. This arrangement 
allowed for multi-hits in the same telescope to be identified and thereby dramatically increase the detection efficiency.

The $^{13}$O secondary beam of $E/A$=65.4~MeV had an intensity of $\approx$10$^5$ pps and a purity of 80\% with $^{11}$C being the major contaminant. More details can be obtained in Refs.~\cite{Webb:2019,Webb:2019a,Webb:2020,Charity:2020,Charity:2021}. The $^9$C secondary beam of $E/A$=68~MeV had an intensity of  1.2$\times$10$^5$ and a purity of $\approx$50\%, with the $^6$Li being the major contaminant and further details can be obtained in Refs.~\cite{Brown:2014a,Brown:2017}

\begin{figure}   
\centering
\includegraphics[scale=0.4]{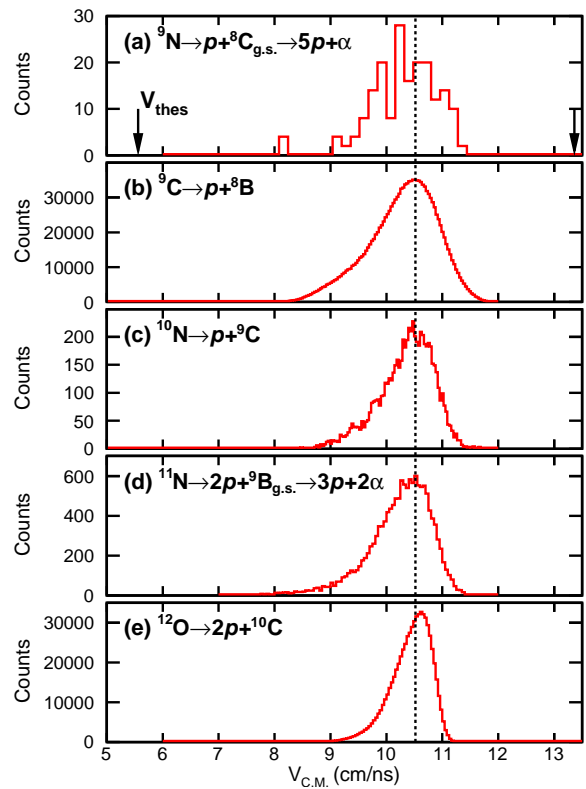}
%home/Boron8/Oxygen11/sortcode_addback/tree/Vcm_all.C 
\caption{\label{fig:Vcm} 
Distribution of reconstructed center-of-mass velocity for some of the resonant-decay channels considered in this work. Panel (c) shows the center-of-mass velocity distribution of the $p$+$^9$C subevents of the detected $d$+$p$+$^9$C events as analysed in Sec.~\ref{sec:N10}. In panel (a), the arrows indicate the low and high-velocity thresholds for proton identification. The dotted vertical lines in all panels give the beam velocity in the center of the target.}
\end{figure}

The angular region subtended by HiRA is these experiments is dominated by projectile fragmentation, a peripheral reaction where the velocities of products are located near to the beam velocity. To illustrate this, Fig.~\ref{fig:Vcm} shows the distribution of center-of-mass velocity of some of the decay channels (produced with the $^{13}$O beam) that are considered in this work. In all cases, these distributions peak close to the beam velocity as indicated by the dotted vertical line.  The widths of these distributions, as expected, steadily increase with increasing mass removed from the projectile \cite{Goldhaber:1974}. These distributions are not influenced by the detection thresholds. For instance in Fig.~\ref{fig:Vcm}(a) for the 5$p$+$\alpha$ channel, the arrows mark the low and high-velocity thresholds for proton identification. The $\alpha$ particle also has the same low-velocity threshold as the proton and its high-velocity threshold is much larger, off-scale in this figure. 

High and low-velocity thresholds for other decay products are also not of concern except for $d$ and $t$ fragments which 
punch though the $\approx$4-cm-thick CsI(Tl) crystals at $E/A$=62 and 57 MeV, respectively. The high-energy threshold for deuterons is about the same as the beam $E/A$ and thus we were not be able to identify roughly 1/2 of the deuterons from projectile fragmentation. We lose even a larger faction of the tritons. 

In the region $\theta_{lab} < 12^\circ$ probed by the HiRA array, a proton-detection efficiency with typical angular distributions is very close to $p$=1/3, thus the probability of not detecting it, 1-$p$, occurs with about twice the probability. In this work, efficiencies, the experimental invariant-mass resolution, and detector biases are based on Monte Carlo simulations \cite{Charity:2019} and these are utilized in the fitting procedure as was done in our previous work.

Instead of the invariant mass $M_{inv}$, results will be presented is terms of the quantity $E_{T}$ which is the $M_{inv}$ minus the sum of the ground-state masses of the decay products and represents the total decay energy of a resonance.  Alternatively if the ground-state mass 
of the nucleus in which resonance resides is well determined,  we will present results in terms of the excitation energy $E^*$ relative to this mass.

\section{excited states in $^9$C}
\label{sec:C9}
We start our investigation of  ``dirty'' projectile-breakup processes by considering $^9$C resonances that decay by 1$p$  emission and are produced by the direct removal of two protons and two neutrons, in total, from the projectile. These removed nucleons can be free or bound in clusters.
With the possibly of two free prompt protons, this is a good test case to see the extent of their background contribution.  The $^9$C resonances are also good test cases as the yields are quite high. In addition, the residual after proton decay ($^8$B) can be selected very cleanly in the $E$-$\Delta E$ spectrum as the neighboring isotopes are particle unstable and thus are missing from this spectrum. 
For some resonances produced in heavier nuclei in this experiment, small background contributions, where the 
residuals was misidentified as a neighbouring isotope, had to be subtracted \cite{Webb:2019,Charity:2021b}. However for the $^9$C resonances, this is not the case.

The nucleus $^{9}$C has one and two proton separation energies of $S_{p}$=1.511 MeV and $S_{2p}$=2.111~MeV \cite{AME2020}. With these two values being so close, we do not expect a large number of excited states that decay by only 1$p$ emission. Thus only a few low-lying states are expected in this channel and as $^8$B has no particle-bound excited states, we only need to consider $p$+$^8$B(g.s.) resonances making the interpretation of this channel simpler than many others.

\begin{figure}
 \centering
\includegraphics[scale=0.42]{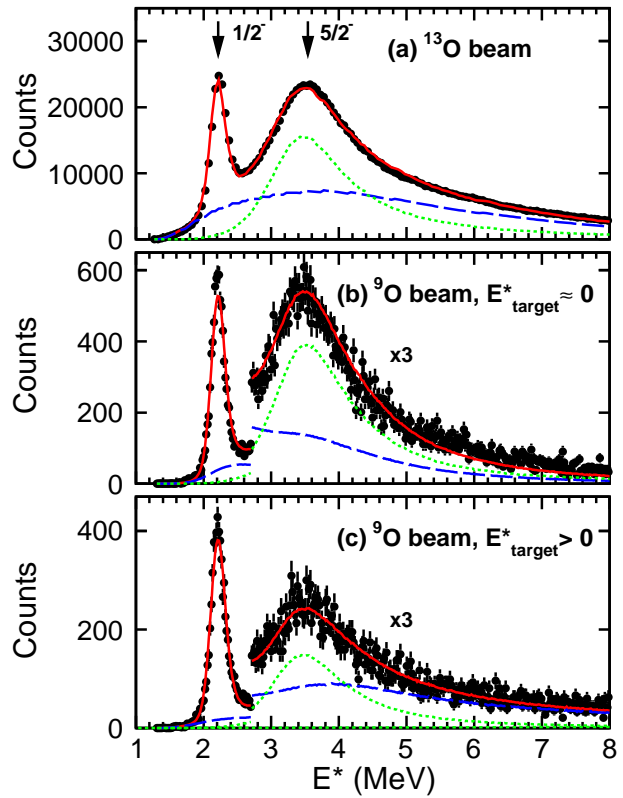}
%home/Boron8/Oxygen11/sortcode_addback/tree/C9/pB9/fit.C 
\caption{\label{fig:pB8} Distribution of $^9$C excitation energy derived from the invariant masses of detected $p$+$^8$B events.  Experimental data points were obtained with (a) the $^{13}$O beam and (b, c) with the $^9$C beam. The latter are subdivided in events where (a) the target is not excited and (b) all other events. Solid red curves show fits to these distributions with backgrounds from the prompt protons indicated with the  dashed blue curves and the contribution from the wide 5/2$^-$ state indicated by the green dotted curves. The results for the $^9$C beam in (b, c) are scaled by a factor of three above $E^*$=2.7 MeV.}
\end{figure}

The $E^*$ spectrum obtained from detected $p$+$^{8}$B events with the $^{13}$O beam is shown by the data points in Fig.~\ref{fig:pB8}(a) and already one can clearly see two known $^9$C excited states. A narrow ($\Gamma$=52~keV) $J^\pi$=1/2$^-$ state state at $E_{T}$=2.218-MeV \cite{ENSDF}, and above this, a very wide $J^\pi$=5/2$^-$ state.  Clearly this spectrum is not overwhelmed from the background due to the two prompt protons which suggests these protons are typically emitted to larger angles and/or are bound in clusters. 

There are two conflicting measurements of the width of the 5/2$^-$ state; one from $p$+$^8$B elastic scattering \cite{Hooker:2019} giving $\Gamma$=1.1(3)~MeV and another from an invariant-mass measurement following the inelastic scattering of a $^9$C projectile \cite{Brown:2017} giving $\Gamma$=0.673(50)~MeV . The inelastic-scattering reaction may also naively be  considered as ``clean'' with no promptly emitted  protons to provide a background to the invariant-mass spectrum. However like the 1$n$ and 2$n$ knockout reactions, ``dirty'' processes will provide a background in the $p$+$^8$B channel, namely from 1$p$ knockout reactions where both the knocked-out proton and the $^8$B residual are detected. In extracting the intrinsic width from the invariant-mass spectrum in Ref.~\cite{Brown:2017}, a flat background was assumed.  If a more energy-dependent background was assumed under the peak,  a different  intrinsic width  might be  extracted possibly  resolving this conflict. 

Figures~\ref{fig:pB8}(b) and \ref{fig:pB8}(c) show the $p$+$^8$B
invariant-mass spectrum from complementary subsets of the $^9$C inelastic-scattering data \cite{Brown:2017}. Rather than displaying the spectrum for all events as was shown in Ref.~\cite{Brown:2017}, these have now been subdivided based on the target excitation energy. With of all the projectile nucleons accounted for, the total invariant mass of the target nucleons can be determined from momentum and energy conservation. This distribution has a strong peak at the ground-state mass of the  target nucleus plus a high-energy tail that extends well over 100 MeV above this. Similar distributions were obtained for inelastic scattering of $^7$Be \cite{Charity:2015}, $^{13}$O \cite{Charity:2021a}, and $^{17}$Ne  \cite{Charity:2018} projectiles at similar beam velocities. Figure~\ref{fig:pB8}(b) shows the distribution gated on the events where the target nucleus remains in its ground state while Fig.~\ref{fig:pB8}(c) is for the remaining events. Both of these spectra also show peaks corresponding the  1/2$^-$ and   5/2$^-$ levels with the narrow 1/2$^-$ level  having  relatively larger yields than in the $^{13}$O breakup reaction. Clearly the 5/2$^-$ state sits ontop of a relatively larger background in Fig.~\ref{fig:pB8}(c) compared to Fig.~\ref{fig:pB8}(b) making the latter spectrum more sensitive to its lineshape.

%, i.e.  the peak to background ratio for the 5/2$^-$ level for the nominally ``cleaner'' inelastic-scattering reaction only approaches that of the ``dirty'' $^{13}$O projectile breakup reaction once a gate on the target excitation is applied.  

In the spectrum for the $^{13}$O fragmentation  reaction in Fig.~\ref{fig:pB8}(a), there are a number of processes that can contribute. 
\begin{enumerate}
    \item[(i)] First are events produced in breakup reactions that promptly remove 2$p$+2$n$ from the parent producing a $^9$C resonance that subsequently decays by emitting a proton. The events that we are interested in for invariant-mass analysis are those where just the delayed proton from the decay of the $^9$C resonance is detected. However, we should also have background from events where this proton is not detected, but instead one of the prompt protons from the first step is detected instead.
     \item[(ii)] There are also events where 3 protons and two neutrons are removed in the projectile breakup reaction directly producing $^8$B in its ground state. A coincidence between one of these prompt protons and the residual will contribute to the background.
    \item[(iii)] There can be events where one proton and two neutrons are removed from the projectile leaving a $^{10}$N resonance. Note $^{10}$N is located beyond the proton drip line so even the ground state will proton decay. If the $^{10}$N levels proton decay to either of the 1/2$^{-}$ or 5/2$^{-}$ states in $^9$C, then such events can contribute to the resonance peaks in Fig.~\ref{fig:pB8} if just the second of the two delayed proton is detected. If any of the other protons are detected instead, then this will contribute to the background. 
    \item[(iv)] One can also consider two-neutron knockout events producing highly excited $^{11}$O states which 3$p$ decay to $^8$B. However there is no evidence for such events as we will see.
\end{enumerate}

   In all four processes listed above, three protons are eventually removed from the projectile to leave a $^8$B residual. Detected 2$p$+$^8$B events sample events where two of these protons are emitted in the angular range subtended by the HiRA detectors. Their detected yield is 6.8\% relative to that for the $p$+$^8$B events. Such events will also populate the $p$+$^8$B channel when the extra protons flies into the spaces between the HiRA detectors. Their expected fraction is twice as large, i.e., 6.8\%$\times\frac{1-p}{p}$ = 13.6\%.  The number of 3$p$+$^8$B events was found to be tiny (0.2\% of the $p$+$^{8}$B yield)  and similar arguments can be made to show that there is negligible contribution to the $p$+$^8$B distribution from events where all three protons are emitted into the angular range of the HiRA detectors.
   These arguments again confirm our observation that the prompt protons from the initial breakup step in the reaction are mostly not found as free nucleons  in the angular range probed by the detector. Thus processes (iii) and (iv), associated with $^{10}$N and $^{11}$O resonances,  which have fewer prompt protons should be more strongly represented in the detected 2$p$+$^{8}$B and 3$p$+$^{8}$B events and thus their yields cannot be large in the $p$+$^8$B events. 

\begin{figure}
 \centering
\includegraphics[scale=0.44]{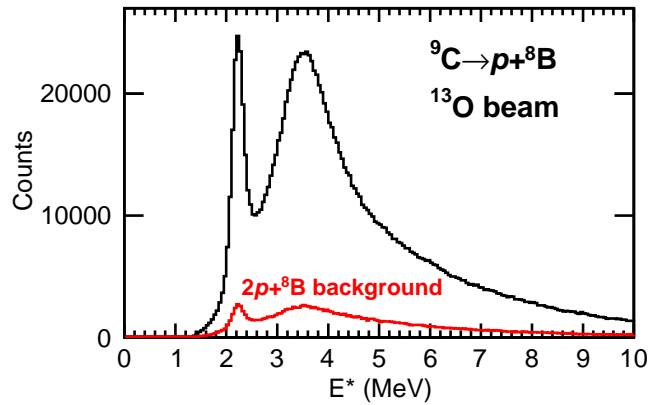}
%home/Boron8/Oxygen11/sortcode_addback/tree/C9/pB9/sub_pB8.C 
\caption{\label{fig:pB8sub} Distribution of $^9$C excitation energy obtained from the invariant masses of detected $p$+$^8$B events using the $^{13}$O-beam data (black histogram). The red histogram shows the contribution from events where two protons and a $^8$B fragment are emitted in the angular region probed with the HiRA array, but where only one of these protons is detected.}

\end{figure}

   Figure~\ref{fig:pB8sub} again shows the $p$+$^8$B invariant-mass spectrum but now compares it to the contribution from events where two protons are emitted in the angular region probed with the HiRA array, but only one is detected. This latter distributions is estimated by the invariant-mass distribution of the two $p$+$^8$B subevents in each 2$p$+$^8$B events. No normalizing is required as this histogram is incremented twice for each 2$p$+$^8$B events giving us the required 13.6\% contribution.  This distribution also displays peaks associated with the two $^9$C proton resonances informing us that  events  from process (i) are significant contributors to the detected 2$p$+$^8$B channel. However there is also a smaller contribution from decay of $^{10}$N states and contributions from process (ii) as will be shown later.

   When both protons and neutrons are removed from the projectile, it is possible that some of these are bound together in clusters. However,
   the proton-rich resonances we are studying decay predominantly by proton emission. Thus clusters observed accompanying such resonances are most likely produced in the initial direct-breakup step of the reaction, i.e. they can be classified as prompt. More care should be taken for $\alpha$ particles as some $\alpha$-cluster states may be expected, but the prompt designation seems especially strong for the neutron-rich deuteron and triton clusters. A study of the correlations between these clusters and the resonances might shed light on the correlations between the prompt protons and the resonances. The larger masses of the clusters suggest that they may be more forward focused than the prompt protons and thus may have larger detection efficiencies.

    We have isolated $cluster$+$p$+$^8$B events and for $d$, $t$, $^3$He, and $\alpha$ clusters they compromise 2.2\%, 1.2\%, 2.9\%, and 11.8\%, respectively, of the detected $p$+$^8$B yield. The yields for $d$ and $t$ clusters are suppressed by their high-energy thresholds. The true number will be roughly twice as large for deuterons (Sec.~\ref{sec:method}) and the factor will be even larger for tritons. If we also account for those clusters that fly in between the HiRA detectors, then $\approx$41\% of $p$+$^8$B events are accompanied by a cluster that is emitted in the angular range probed in this experiment. While the number of clusters emitted at larger angles is unknown, it is clear, for this channel, that a significant number of  promptly removed protons in the first step are not liberated as free protons, but are bound in clusters.

\begin{figure}
 \centering
\includegraphics[scale=0.42]{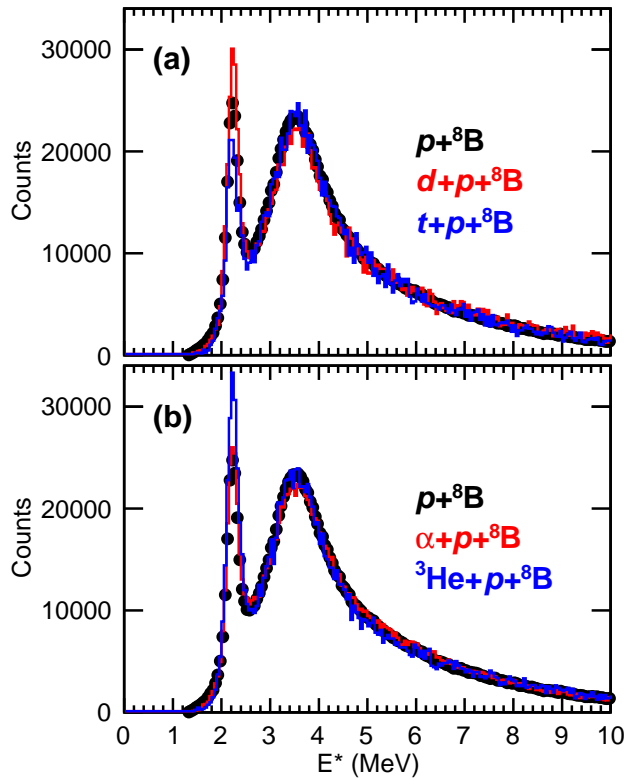}
%home/Boron8/Oxygen11/sortcode_addback/tree/C9/pB9/Erel_dta.C 
\caption{\label{fig:pB8cluster} Comparison of the distributions of $^9$C excitation energy obtained from the invariant mass of detected $p$+$^8$B events (black data points) to those where a cluster was also detected in coincidence (colored histograms). All results were obtained with the ${13}$O-beam data.}

\end{figure}

    The distributions of $^9$C excitation energy determined from the $p$+$^8$B subevents of each $cluster$+p+$^8$B event are compared to those obtained for all detected $p$+$^8$B events in Fig.~\ref{fig:pB8cluster}.
    The distributions have been normalized to the same number of counts in the region above the 1/2$^-$ state to highlight differences in the shape of the 5/2$^-$ resonance and the background underneath it. The distributions show that the presence of a cluster has little effect on the $^9$C excitation-energy distribution, with the largest  difference being small changes in the relative yield of the narrow 1/2$^-$ state. Otherwise all distributions are almost identical in shape. This suggests there are no strong correlations between the promptly emitted clusters and the resonances.

   To further explore the level of these correlations, correlations function have been constructed:
   \begin{equation}
          1 + R(E)   = \frac{F(E)}{F_{mixed}(E)} K
   \end{equation}
   where $F(E)$ is the experimental distribution as a function of some relative energy. For this relative energy, we have considered both 
   $E_{cluster-^9\mathrm{C}}$ and $E_{cluster-^8\mathrm{B}}$, the relative energy between the cluster and the center of mass of the reconstructed $^9$C fragment or between it and the $^8$B residual. The quantity $F_{mixed}$ is the distribution with no correlations between the cluster and the other fragments and it is obtained from event mixing which removes effects from the detector response. We have mixed a cluster from one  $cluster$+$p$+$^8$B event  with  $p$+$^8$B fragments from a different such event which ensures that the correlations between the $p$ and the $^8$B fragment are preserved.   All fragments in a mixed event must be associated with a unique CsI(Tl) crystal as is found in all the real events that are analysed.
   The constant K is adjusted so that the correlation function asymptotically approaches unity with increasing $E$.
   
     \begin{figure}
 \centering
\includegraphics[scale=0.44]{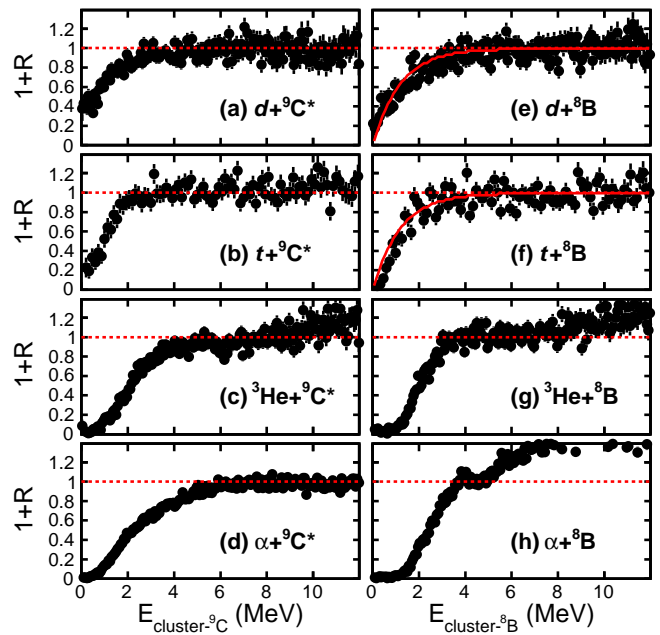}
%home/Boron8/Oxygen11/sortcode_addback/tree/C9/pB9/corr.C 
\caption{\label{fig:corr_pB8} Correlations functions obtained for the $cluster$+$p$+$^8$B events as a function of (a-d) the relative energy between the cluster and the reconstructed $^9$C fragment and (e-h) the relative energy between the cluster and the $^8$B residual. For comparison, the solid red curves in (e, f) are the correlation functions for prompt protons obtained in the fit to the distribution of excitation energy in Fig.~\ref{fig:pB8}(a).
}
\end{figure}

     Correlation functions for the different clusters are shown in Fig.~\ref{fig:corr_pB8}.  They all have similar features with suppressions ($R<$0) for small values of relative energies and asymptoting to $R$=0, i.e. no correlations at larger relative energies. The results for the $\alpha$+$^8$B relative energy in Fig.~\ref{fig:corr_pB8}(h) is a little puzzling as while reaching a plateau of unity at 4 MeV, it subsequently increases at higher relative energies. Perhaps there are contributions from a number of unresolved $^{12}$N$\rightarrow\alpha$+$^8$B resonances in this region.
     
     The suppression feature is at least partially related to long-range Coulomb final-state interactions and presumably is better represented in the $E_{cluster-^{8}\mathrm{B}}$ results of Figs.~\ref{fig:corr_pB8}(e) to \ref{fig:corr_pB8}(h) as most of the $^9$C resonance yield is associated with the very shorted-lived 5/2$^-$ state which breaks up while the other projectile fragments are close by. Indeed the suppression is much larger for the $Z$=2 clusters as compared the $Z$=1 clusters, i.e., compare $^{3}$He+$^8$B to $t$+$^8$B. However, it is not clear that the  Coulomb interaction can explain all the suppression. For instance, the $d$+$^8$B and $^3$He+$^8$B correlations both reach their asymptotic value at   $E_{cluster-^{8}\mathrm{B}} \approx$ 3 MeV even through the Coulomb forces are much larger in the latter case. Possibly the creation of a particular resonance requires that no other projectile nucleons are initially located in neighboring phase-space cells during the first step. 

         \begin{figure}
 \centering
\includegraphics[scale=0.42]{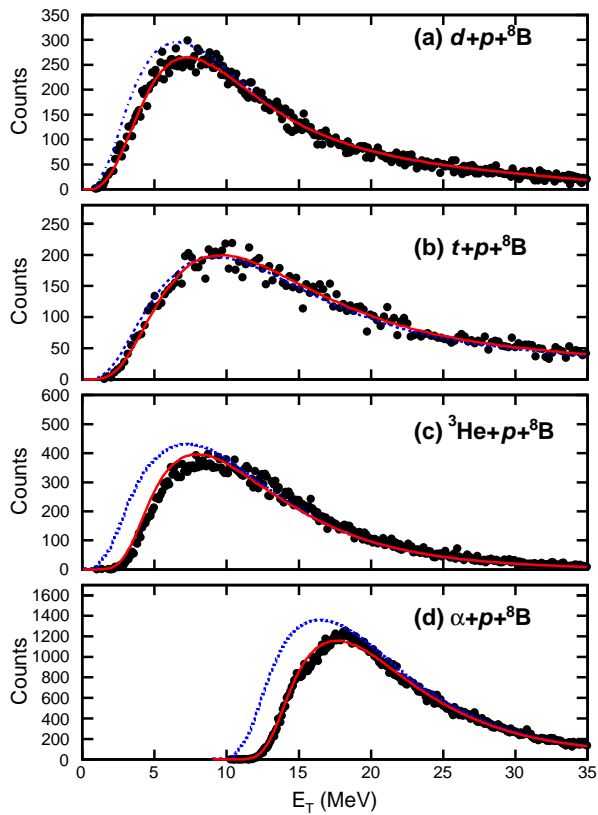}
%home/Boron8/Oxygen11/sortcode_addback/tree/C9/pB9/Erel_dta_3body_core.C 
\caption{\label{fig:Erel_3body} Distributions of total decay energy for $cluster$+$p$+$^8$B events. The data points show the experimental results, the dotted blue curves are from the mixed events, and the solid red curves were obtained when the latter events are weighted by the fitted correlations functions.}
\end{figure}
    
    To further convince ourselves that the clusters are not produced in the decay of $^{11,12}$N or $^{12,13}$O resonances we show the distributions of $E_{T}$ for the $cluster$+$p$+$^8$B events in Fig.~\ref{fig:Erel_3body}.  These distributions show smooth variations with $E_{T}$ with no sharp resonance peaks visible. The dotted blue curves show the results from  uncorrelated mixed events and the solid red curves were obtained when these mixed events are weighted by the correlation functions in Fig.~\ref{fig:corr_pB8}. Almost identical results are obtained if we used the E$_{cluster-^9\mathrm{C}}$ or $E_{cluster-^8\mathrm{B}}$ correlations for weighting. These weighted mixed events reproduce the experimental results to a high degree and confirm that clusters have minimal yield from resonant decay.

    With the clusters showing little correlations with the resonances and their decay products, apart from a suppression when they are initially close in phase space,
    it is natural to assume that the prompt protons will also follow the same trend. However we are left with the question as to whether the suppression for these prompt protons is similar to the other hydrogen isotopes or not? To explore this we have fit the excitation-energy distribution of Fig.~\ref{fig:pB8}(a)  with two resonance peaks and a background from prompt protons. The shape of the background distribution is obtained from mixing $p$ and $^8$B particles from different $p$+$^8$B events and weighting these mixed events with a modified inverse-Fermi function
\begin{multline}
     1 + R(E) = \exp \left(-\frac{c_1}{c_2}\right) \\
     \times \left[ \frac{1+\exp(c_{1}/c_{2})} {1+\exp(-\frac{E-c_1}{c_2})}-1 \right] \label{eq:supp}
\end{multline}
    where $c_1$ and $c_2$ are two fit parameters and $E$ in this case is the $p$+$^8$B relative energy. This parameterization is able to fit the suppressions in all the correlation functions of Fig.~\ref{fig:corr_pB8} quite well. The  intrinsic lineshape for the narrow 1/2$^-$ resonance is taken as a Breit-Wigner form with the centroid and width constrained to  its tabulated value \cite{ENSDF}. The less establish lineshape of the wide 5/2$^-$ resonance is allowed to vary and parameterized as an isolated resonance in $R$-matrix theory \cite{Lane:1958} for $\ell$=1 with a channel radius of 4.2~fm. The effect of the detector resolution and efficiency on the lineshapes was taken into account from Monte Carlo simulations \cite{Charity:2019}. There are 7 fit parameters; three associated with the magnitudes of the two resonances and the background, two associated with Eq.~(\ref{eq:supp}), and other two from the line shape of the 5/2$^-$ resonance. 
    
    The fitted distribution is shown as the solid red curve in Fig.~\ref{fig:pB8}(a) where the fitted background is indicated as the dashed blue curve with fit parameters of $c_1$=-38.4~MeV and $c_2$=1.19~MeV.  In Figs.~\ref{fig:corr_pB8}(e) and \ref{fig:corr_pB8}(f), the fitted correlation function for prompt protons (solid red lines) is compared to those obtained for the other hydrogen isotopes. The correlation functions for all hydrogen isotopes are very similar suggesting that those for $d$ and $t$ clusters can provide a reliable prediction for the prompt protons.
    
    In terms of the lineshape of the 5/2$^-$ resonance, we have looked for consistency with the $^9$C-beam data. 
    Mixed events from these data sets were weighted with the same correlation function as used in the fit to the $^{13}$O-beam data  in Fig.~\ref{fig:pB8}(a). Similarly the line shape of the 5/2$^-$ state was fixed to the result from that previous fit as well. The red curves in Figs.~\ref{fig:pB8}(b) and \ref{fig:pB8}(c) show the results of these fits where only the magnitudes of the three components were allowed to vary. This prescription also allows these $^9$C-beam data to be well reproduced. The fitted FWHM of the 5/2$^-$ resonance is  1.33(13)~MeV. This is consistent with the value of 1.1(3)~MeV from elastic proton scattering on $^8$B in \cite{Hooker:2019}. The disagreement between this $p$+$^8$B elastic-scattering study and the $^9$C-beam data is now resolved in favor of the result from the former.

\begin{figure}
 \centering
\includegraphics[scale=0.42]{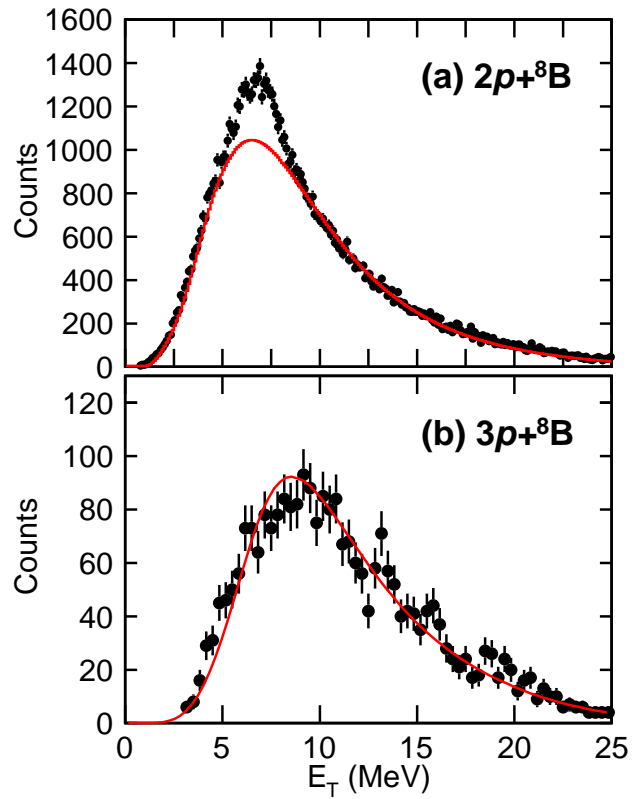}
%home/Boron8/Oxygen11/sortcode_addback/tree/C9/pB9/Erel_10N_11O.C 
\caption{\label{fig:Erel_10N} Distributions of total decay energy for (a) 2$p$+$^8$B and (b) 3$p$+$^8$B events. Data points are the experimental results. The solid red curves show the expected background from $p$+$^8$B and 2$p$+$^8$B events, respectively,  where an extra prompt proton is included. }
\end{figure}

    Finally, it is of interest to determine the relative intensities of processes (iii) and (iv) associated with the decay of $^{10}$N and $^{11}$O resonances.  The $E_{T}$ distributions obtained for 2$p$+$^8$B and 3$p$+$^8$B events, shown in Fig.~\ref{fig:Erel_10N}, have no narrow resonance  features. To search for contributions from wider resonances,  we have constructed the background distributions from coincidences with prompt protons shown as the solid red curves. For $^{10}$N, detected $p$+$^8$N events were mixed with a proton from another such event.  These mixed events were weighted  by our fitted prompt-proton correlation function using the relative energy of the proton and $^8$B fragments from different events. For  $^{11}$O, the background was obtained in a similar manner starting from the detect 2$p$+$^8$B events and mixing in another proton. 
    
    The result for $^{11}$O, shown in Fig.~\ref{fig:Erel_10N}(b), reproduces the experimental data reasonable well. Thus this distribution can be understood from the 2$p$+$^8$B events by including another prompt proton and there is no unexplained yield that could be attributed to wide $^{11}$O resonances. However for $^{10}$N in Fig.~\ref{fig:Erel_10N}(a), while the high and low-$E_{T}$ tails can be reproduced by our background, these is excess yield at $E_{T}\sim$5~MeV that must be due to contributions from one or more $^{10}$N excited states.  The excess yield accounts for 10\% of the total and thus the contribution from $^{10}$N resonant decay to the detected $p$+$^8$B yield is 10\% $\times$ 13.6\% = 1.36\%  which can be ignored. If these $^{10}$N yield was more significant, one could  subtract its 
    contribution from the $p$+$^8$B distribution before fitting.
    
  In summary for this channel, the yield of $p$+$^8$B, 2$p$+$^8$B, 3$p$+$^8$B, and $cluster$+$p$+$^8$B events are dominated by processes (i) and (ii), i.e.  $^8$B residuals are produced  either directly in the first step or via decay of $^9$C resonances. In our fit to the $p$+$^8$B invariant-mass distribution  in Fig.~\ref{fig:pB8}(a), 45\% of the yield was from the former. 

\begin{figure}
 \centering
\includegraphics[scale=0.45]{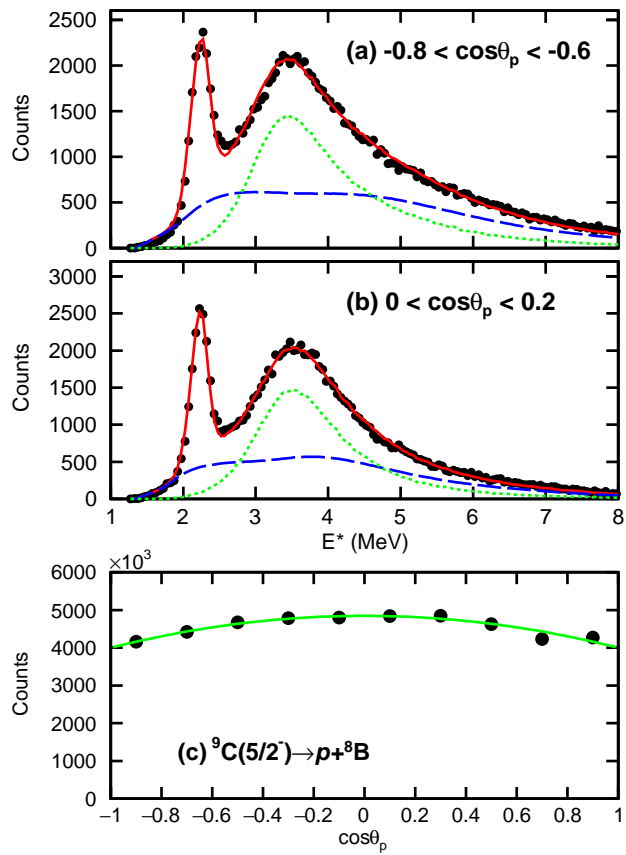}
%home/Boron8/Oxygen11/sortcode_addback/tree/C9/pB9/fit_c/ang.C 
\caption{\label{fig:ang_pB8} (a, b) Distributions of $^9$C excitation energy gated on the indicated regions of proton decay angle $\theta_{p}$.  The solid red curves are fits to these distributions including the prompt proton background distributions (dashed blue curves) obtained from event mixing with the same $\theta_{p}$ gates.  The dotted green curves give the fitted contribution from the 5/2$^-$  state.  (c) The angular distribution obtained from these fits  where the error bars are smaller than the size of the data points. The solid green curve shows a fit using Eq.~(\ref{eq:ang})}.
\end{figure}

  It is often useful to consider invariant-mass distributions gated on the decay angle of the proton $\theta_{p}$ in the parent's reference frame where $\theta_{p}$=0 corresponds to the beam axis. Such gates can be used to determine the decay angular distribution  which can yield information on the spin/parity and structure of the state.  Also some regions of $\theta_{p}$ have better invariant-mass resolution than others \cite{Charity:2019,Charity:2019a} and so such gating can be useful to resolve fine structures in invariant-mass distributions.  We have therefore fit the $p$+$^8$B invariant-mass distribution gated on $\theta_p$. To be consistent, the same $\theta_p$ gate was applied to both the real and mixed events. Figures~\ref{fig:ang_pB8}(a) and \ref{fig:ang_pB8}(b) show some typical examples of the fitted distributions in these cases for  -0.8$<\cos\theta_p<$-0.6 and 0$<\cos\theta_p<$0.2 gates, respectively. One can see small modifications in the shape of both the experimental and background distributions between the different $\theta_p$ regions, but the quality of the fits is always excellent. 
  
  One test of this procedure is that the angular distribution of the  5/2$^-$ state above background obtained from these gates should be realistic, i.e. symmetric about $\cos\theta_p$=0.  This angular distribution is shown in Fig.~\ref{fig:ang_pB8}(c) and indeed this symmetry is present. However it is also clear that the measured angular distribution has a small anisotropy, i.e. it is not flat. The observation of this anisotropy implies that there is some level of spin alignment produced in this projectile breakup reaction. Indeed spin alignment and polarization effects have been observed in such reactions before \cite{Ichikawa:2012}.  For such a light nucleus, the proton decay of the 5/2$^-$ state in $^{9}$C to the 2$^+$ ground state of $^8$B should be dominated by $p$-wave emission. The emission of a $p_{1/2}$ proton is isotropic, so the  observed anisotropy must indicate some level of $p_{3/2}$ emission. For pure $p_{3/2}$ emission or mixed $p_{1/2}$-$p_{3/2}$ emission where there is an interference term, the angular distribution will have contributions from the square of $\ell$=1 Legendre polynomials with just $|m|$=0,1 and thus will have the general form 
  \begin{equation}
  \label{eq:ang}
  d\sigma/d\cos\theta_{p} = K_1 + K_2 \cos^{2}\theta_p,
\end{equation}
with only two unknown constants ($K_1$ and $K_2$). The solid curve in Fig.~\ref{fig:ang_pB8}(c) shows a fit with this form which allows an excellent reproduction of  the experimental angular distribution.

   \section{Levels in $^{10}$C}
   The adjacent isotope to $^9$C, $^{10}$C has only one particle-bound excited state (2$^+_1$) and above that most levels decay into the 2$p$+2$\alpha$ exit channel \cite{ENSDF}.  In making such $^{10}$C states from the breakup of the $^{13}$O projectile, there is the possibility of 2 prompt protons in addition to the two from resonant decay. Hence there are potentially many different ways of producing backgrounds under the $^{10}$C resonance peaks.   Nevertheless, after a careful analysis, data from the present $^{13}$O experiment and other experiments yielded a consistent level structure for $^{10}$C \cite{Charity:2022}. 

    The complexity of problem can be simplified if one can gate on a narrow intermediate level in the sequential decay of these $^{10}$C states. In this work, we will reconsider the case for decays with the $^9$B ground-state ($\Gamma$=0.54~keV) as the intermediate state and concentrate on the backgrounds associated with prompt protons for this channel. As shown in Ref.~\cite{Charity:2022}, the invariant-mass distribution of the two $p$+2$\alpha$ subevents in each 2$p$+2$\alpha$ event shows a strong narrow peak just above threshold which sits on negligible background.  By gating on this peak, events with a $^9$B(g.s.) intermediate fragment can be cleanly isolated and the proton from the decay of the $^{10}$C resonance to $^9$B(g.s.) can be identified. We can thus consider these events as the two-body channel $p$+$^9$B(g.s.)  which makes the discussion of the background very similar to the $p$+$^8$B events in the preceding section.

\begin{figure}
 \centering
\includegraphics[scale=0.45]{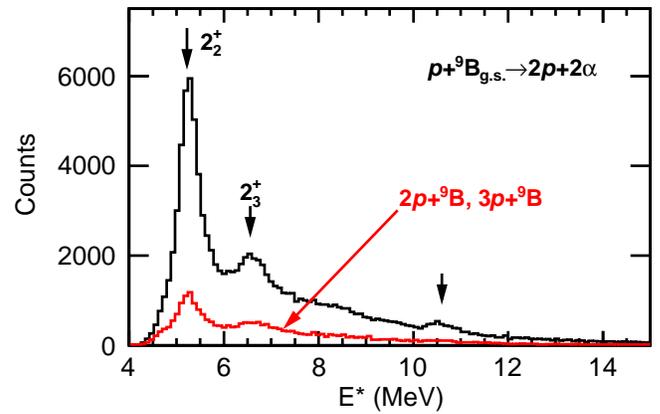}
%home/Boron8/Oxygen11/sortcode_addback/tree/C10/2p2a/Ex_sub.C 
\caption{\label{fig:pB9} Distributions of $^{10}$C excitation energy 
obtained from  the detected $p$+$^9$B(g.s.)$\rightarrow$2$p$+2$\alpha$ events. The background from 2$p$+$^9$B(g.s.) and 3$p$+$^9$B(g.s.) events where the  protons are emitted in the angular range probed by the HiRA array, but only one is detected, is shown as the red histogram. Arrows show the location of $^{10}$C states known to decay into this channel.}
\end{figure}

    The $^{10}$C excitation energy obtained from these $p$+$^9$B(g.s.) events is shown in Fig.~\ref{fig:pB9} as the black histogram. Three 
    resonances are visible whose centroids, indicated by the arrows, have been constrained in previous studies \cite{ENSDF,Charity:2022}. These peaks are relatively narrow and there are regions where the distribution is free of resonances and thus dominated by the background from prompt protons.
    
    Previous analyses of these data has also indicated that there are levels in $^{11}$N and $^{12}$O 
     which decay via sequential two and three-proton emission to 
     $^9$B(g.s.) \cite{Webb:2019a,Charity:2022} and thus will contribute to the detected $p$+$^9$B(g.s.) events. The correlations of these protons with the $^9$B(g.s.) intermediate state will be quite different from the correlations associated with the prompt protons and thus their contributions must be accounted for separately.   
     
     The red histogram in Fig.~\ref{fig:pB9} shows the distributions of events where two and three protons are emitted into the angular region probed by the HiRA detectors, but only one of the protons is detected. This contribution is determined from the detected 2$p$+$^9$B(g.s.) and 3$p$+$^9$B(g.s.) events and contains most of the contributions from processes (iii) and (iv) (i.e. $^{11}$N and $^{12}$O decay), but will also contain significant number of events from the other  two processes as well. The red histogram accounts for 24\% of the detected $p$+$^9$B(g.s.) events which is about twice as much as found in the equivalent distribution for the $p$+$^8$B events in Fig.~\ref{fig:pB8sub}. As we will see later, this larger fraction is predominantly due to the contributions from the decay of $^{11}$N states. This component is subtracted from the spectrum of detected $p$+$^9$B(g.s.) events (black histogram) is order to remove all contributions from $^{11}$N and $^{12}$O decay. This subtraction is excessive in the sense that only a fraction of the 2$p$+$^9$B(g.s.) and 3$p$+$^9$B(g.s.) are from decay of higher-Z resonances. However it does ensure that the contamination from higher-Z resonances is removed.

\begin{figure}
 \centering
\includegraphics[scale=0.45]{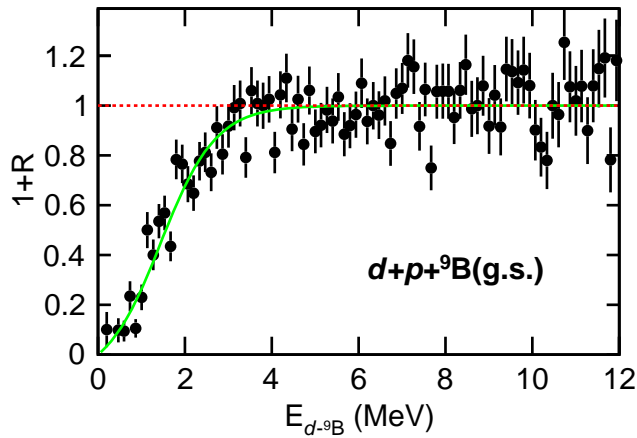}
%home/Boron8/Oxygen11/sortcode_addback/tree/C10/2p2a/corr.C 
\caption{\label{fig:corr_pB9} Correlation function obtained from the $d$+$p$+$^9$B(g.s.)$\rightarrow d$+2$p$+2$\alpha$ events
as a function of the relative energy between the $d$ and the reconstructed $^9$B fragment.  The solid green curve shows a fit with Eq.~(\ref{eq:supp}).}

\end{figure}

     Events with clusters accompanying the $p$+$^9$B(g.s.) events were also searched for. However, only the $d$+$p$+$^9$B(g.s.) channel had sufficient statistics to construct a correlation function which is plotted in Fig.~\ref{fig:corr_pB9}.  The correlation here is very similar to that obtained for the $d$+$^8$B events in Fig.~\ref{fig:corr_pB8}(e).  The solid green curve is a fit with Eq.~(\ref{eq:supp}) and is used to weight mixed events in order to construct the background from coincident prompt protons. 

\begin{figure}
 \centering
\includegraphics[scale=0.45]{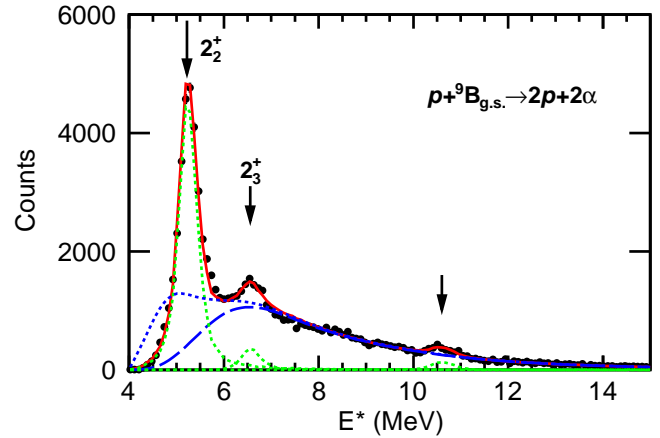}
%home/Boron8/Oxygen11/sortcode_addback/tree/C10/2p2a/Ex_fit.C 
\caption{\label{fig:fit_pB9} 
Data points show the $^{10}$C excitation-energy distribution for the $p$+$^9$B(g.s.)$\rightarrow$2$p$+2$\alpha$ events corrected for the background from 2$p$+$^9$B(g.s.) and 3$p$+$^9$B(g.s.) events. The solid red curve shows the fit to these data with three $^{10}$C resonances and the background (dashed blue curve) associated with detecting prompt protons.
The dotted green curves show the contributions for each resonance.
The dotted blue curve, which was obtained from the mixed events without weighting and normalised to reproduce the high-energy tail, cannot reproduce the data at low values of $E^*$.}

\end{figure}

     Figure~\ref{fig:fit_pB9} shows a fit to the invariant-mass data using this background (dashed blue curve). The data points in this plot are the distribution where the contributions from the 2$p$+$^9$B(g.s.) and 3$p$+$^9$B(g.s.) events (red histogram in Fig.~\ref{fig:pB9}) have been subtracted to remove  contamination from $^{11}$N and $^{12}$O decay.
     The resonance parameters for the two lowest-energy resonances (2$^+_2$ and 2$^+_3$) were fixed to values from a previous study using inelastic scattering \cite{Charity:2009} while the parameters for the highest-energy resonance were obtained from a previous analysis of these data. Only the magnitudes of the three resonance and the background were adjusted in the fit which reproduced the data exceedingly well. This provides confidence in our background prescription.
     
     We also note that weighting of the mixed events is essential in order to get an acceptable reproduction of these data. The dotted blue curve in this figure was obtained from the mixed events without weighting and normalized to reproduce the high-energy tail of the experimental data. This unweighted background must be rejected as it gives too much yield at the smallest excitation energies.

\begin{figure}   
\centering
\includegraphics[scale=0.42]{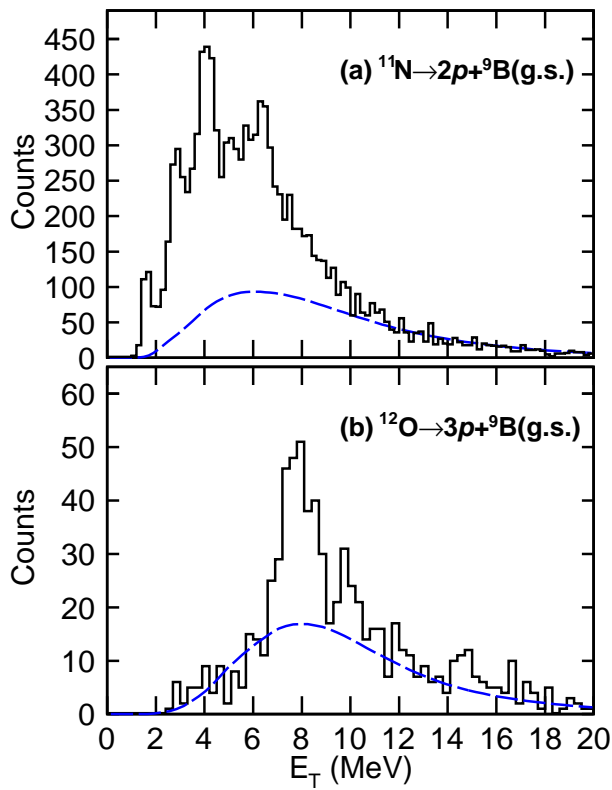}
%home/Boron8/Oxygen11/sortcode_addback/tree/C10/2p2a/Erel_11N_12O.C 
\caption{\label{fig:Erel_11N_12O} 
Data points show the decay-energy spectra for (a) $^{11}$N$\rightarrow$2$p$+$^9$B(g.s.)$\rightarrow$3$p$+2$\alpha$ events and (b) $^{12}$O$\rightarrow$3$p$+$^9$B(g.s.)$\rightarrow$4$p$+2$\alpha$ events. The distribution in (a) has been corrected for the contribution 3$p$+$^9$B(g.s.) events where one of the protons flies in between HiRA telescopes and is not detected. The dashed blue curves show the predicted backgrounds from coincidences with prompt protons estimated using weighted mixed events.}

\end{figure}

The $E_{T}$ distributions for 2$p$+$^9$B(g.s.) and 3$p$+$^9$B(g.s.) events are shown in Fig.~\ref{fig:Erel_11N_12O} as the histograms.  The results for 2$p$+$^9$B(g.s.) events have been decreased by the contribution from events where there are three protons in the angular range spanned by the HiRA array, but only two are detected to remove contamination from $^{12}$O decay.  Clearly resonance peaks associated with $^{11}$N and $^{12}$O states are visible in these distributions.

The dashed blue curves show the backgrounds from coincident prompt protons. For the $^{11}$N states, this was obtained from mixing an additional proton with detected $p$+$^9$B(g.s.) events and weighting these by the fitted correlation function in Fig.~\ref{fig:corr_pB9}. Similar to the $^{12}$O resonances, we have mixed an additional proton into the detected 2$p$+$^9$B(g.s.) events.  These backgrounds have been normalized to account for the tail regions of the experimental spectra.  While the $^{11}$N distribution in Fig.~\ref{fig:Erel_11N_12O} has some narrow resonances, there must also be wider unresolved resonances present  to explain all the yield above the prompt proton background. From the  yields above background in these plots, we estimate the detected $p$+$^9$B(g.s.) events have a 1.7\% contribution from $^{12}$O resonances produced in 1$n$ knockout reactions and a 21\% contribution from $^{11}$N resonances produced after the removal of 1$n$ and 1$p$ from the projectile. From the fit in Fig.~\ref{fig:fit_pB9}, we estimate that 46\% of the detected $p$+$^9$B(g.s.) events are background associated with prompt protons and where the $^9$B(g.s.) resonance was created in the first step of the reaction.

\section{Levels in $^{11}$N}
\label{sec:N11}
We have just discussed $^{11}$N levels that 2$p$ decay to $^{9}$B. In this section we will concentrate on the $p$+$^{10}$C decay channel for which  an analysis of the present data has been presented in Ref.~\cite{Webb:2019a}. Here we will reconsider this channel with our new prescription for the background. The creation of a $^{10}$C fragment involves the removal of 2 protons and a neutron, in total, from the projectile.    With fewer total number of protons involved, the number of ways of producing a background is reduced.  

\begin{figure}   
\centering
\includegraphics[scale=0.53]{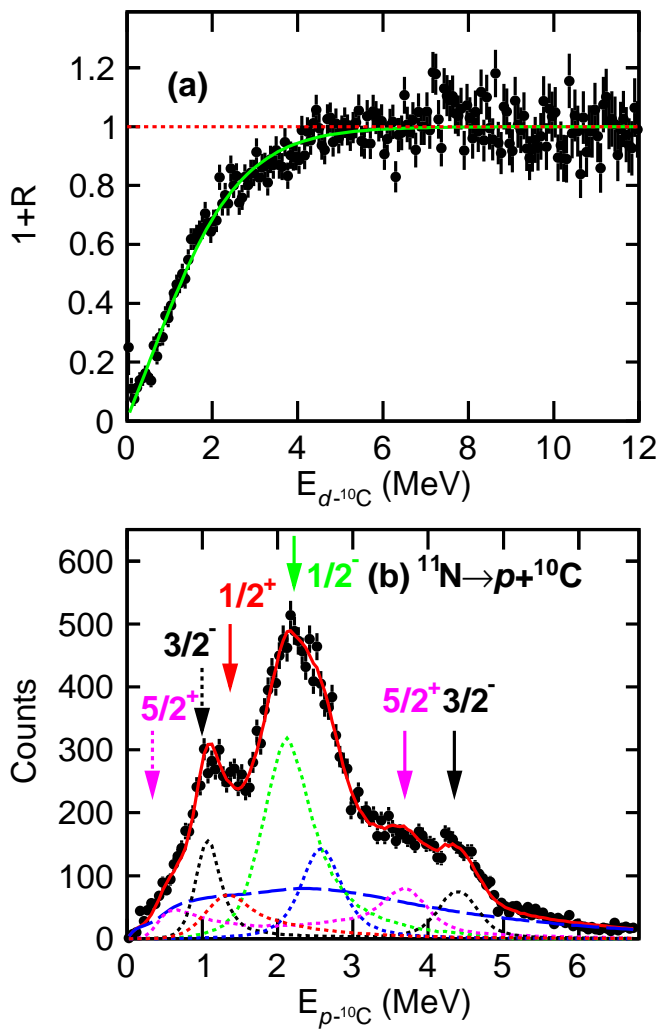}
%home/Boron8/Oxygen11/sortcode_addback/tree/N11/pC10/N11.C 
\caption{\label{fig:N11} 
Data obtained from $d$+$p$+$^{10}$C events.
The $d$-$^{10}$C correlation function is shown in (a) plotted verses their relative energy. The distribution of $^{11}$N decay energy obtained from the $p$+$^{10}$C subevents is shown in (b) with the results of the fit shown by the solid red curve. The fitted background from the prompt protons is indicated by the dashed blue curve, while the contributions from the different $^{11}$N states are indicated by the dotted curves. Solid arrows indicate the location of previously known states, the dashed arrows locate peaks associated with proton decay the the 2$^+$ first excited state of $^{10}$C as assumed in the fit.}
\end{figure}

In the previous analysis of this $p$+$^{10}$C data, a 37\% correction was made for 2$p$+$^{10}$C events where one of the protons flew in between the HiRA detectors. A significant fraction  of these events are from the prompt 2$p$ decay of $^{12}$O  states which do not decay through $^{11}$N intermediate states. An additional few percent was also subtracted due to  $p$+$^{11}$C events for which the $^{11}$C fragment was misidentified as a  $^{10}$C fragment. In order to avoid these subtractions which may introduce same systematic uncertainty, we now propose an alternative method.  To the extent that accompanying clusters do not bias the invariant-mass distribution for proton decay as demonstrated in $^9$C$\rightarrow p$+$^8$B decay (Sec.~\ref{sec:C9}),
then proton decay of $^{11}$N can be studied from $d$+$p$+$^{10}$C events. The presence of the deuteron precludes the possibility that a $^{12}$O state was formed in the event. In addition, contamination from misidentified $d$+$p$+$^{11}$C events is suppressed as this would involve the pickup of a neutron from the target nucleus which is expected to have a small cross section. Figure~\ref{fig:N11}(b) shows the $^{11}$N decay-energy spectrum obtained from the $p$+$^{10}$C subevents. To enable comparison with the original analysis, we have gated on $|\cos\theta_p|<$0.5 which improves the energy resolution.

The shape of the spectrum is quite similar to the original contamination-subtracted spectrum in Ref.~\cite{Webb:2019a}.  The largest difference, is at small values of $E_{T}$, where the misidentified $^{11}$C component was largest.
The downside to this new method is the considerable reduction in statistics with the required deuteron coincidence.

In order to determine the background from prompt protons, we have used the $d$+$^{10}$C subevents to construct the correlation function of Fig.~\ref{fig:N11}(a). This correlation function is very similar to $d$+$^{8,9}$B functions shown in Figs.~\ref{fig:corr_pB8}(e) and \ref{fig:corr_pB9}. 
Figure~\ref{fig:N11}(b) shows a possible fit to the $p$+$^{10}$C decay-energy spectrum using this background.  The result of this fit is similar to that made in the original analysis. However in this previous work, the shape of the background was  allowed to vary with a simple prescription that permitted only smooth variations with $E_{T}$. 
In the present analysis, the shape of the background is prescribed and its magnitude is largely determined from fitting the  region above $E_{T}$=5~MeV  where it dominates the spectrum. 

Because $^{10}$C has one particle-stable excited state ($J^\pi$=2$^+$, $E^*$=3.35~MeV) which $\gamma$ decays, there are contribution from $^{11}$N states that proton decay to both the ground and to this excited state. The decay energy for the latter are shifted down by the excitation energy of this excited state.  Unfortunately none of the peaks associated with these various states are well resolved. In the fit of this and the previous analysis, peaks for the known  1/2$^-$, 3/2$^-$ and 5/2$^+$  states are constrained with their evaluated centroids [solid arrows in Fig.~\ref{fig:N11}(b)] and intrinsic widths \cite{ENSDF}.  Between the 1/2$^-$ and 3/2$^+$ resonance, more strength was needed so another peak was introduced. In the original analysis, this peak has the parameters of $E_{T}$=2.563(10)~MeV and $\Gamma$=697(32)~keV and in the present analysis these are now $E_{T}$=2.600(13)~MeV and $\Gamma$=111(30)~keV. The shifts in these quantities beyond the listed statistical uncertainties are due to the improved analysis.  Possibly this new state is associated with decay to the 2$^+$ excited state in $^{10}$C as it was not seen in $p$+$^{10}$C elastic scattering \cite{Axelsson:1996,Markenroth:2000,Casarejos:2006}.   The lower-energy region below $E_{T}$=1.5~MeV cannot be explained by the decay of the ground state of $^{11}$N (red dotted curve), instead it appears to be dominated by proton decays to the 2$^+_1$ excited state of $^{10}$C. In the original analysis, the peak at $E_T\approx$1~MeV was associated with a decay branch of the 3/2$^-$ state to this excited state. In the new analysis, more yield at even lower energies is required and so a similar decay branch of the 5/2$^+$ state is invoked to fit the spectrum.  While the case for the branch of the 3/2$^-$ state seems robust, the case for the 5/2$^+$ state will depend very much on our prescription for the prompt proton background. Future studies should verify this branch. Finally, the fraction of events in the fitted spectrum [Fig.~\ref{fig:N11}(b)] from the prompt protons is 29\%.

\section{Levels in $^{10}$N}
\label{sec:N10}
Detected $p$+$^9$C events from the $^{13}$O-beam data have been previously analysed in Ref.~\cite{Charity:2021b} in order to investigate $^{10}$N levels. Before this, only two previous studies had been made of this isotope \cite{Lepine:2002,Hooker:2017}.
In our previous analysis, before fitting the $p$+$^9$C spectrum, a 21\% contribution from 2$p$+$^9$C events where one of the protons flies between the HiRA detector was subtracted. This subtractions removes contributions from $^{11}$O$\rightarrow$2$p$+$^9$C decays \cite{Webb:2019}.
In addition a 10\% contamination from $p$+$^{10,11}$C events where the carbon isotopes were misidentified as a $^9$C fragment
was also subtracted. 

\begin{figure}   
\centering
\includegraphics[scale=0.45]{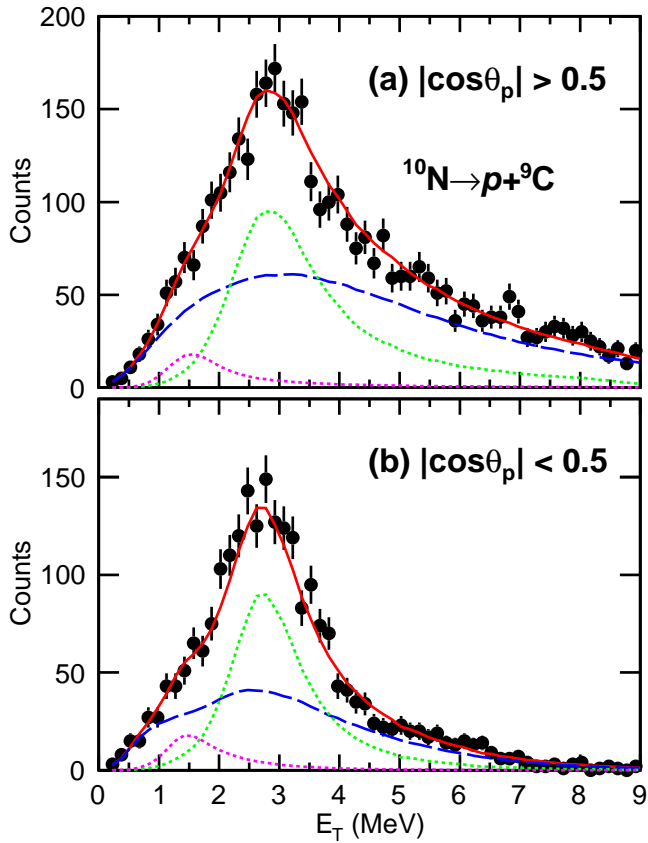}
%home/Boron8/Oxygen11/sortcode_addback/tree/N10/pC9/fitBackTriton_tl/fit_N10.C 
\caption{\label{fig:N10} 
Experimental distribution of $^{10}$N decay energy obtained from the $p$+$^9$C subevents of the detected $t$+$p$+$^9$C events. The distributions have been gated on (a) longitudinal decay $|\cos\theta_p| > 0.5$ and (b) transverse decay $|\cos\theta_p|<0.5$. The solid red curves show a joint fit to these data with two peaks (dotted curves) and the background from prompt protons (dashed blue curves). }
\end{figure}

In the present analysis, we will again use the alternative method to avoid these backgrounds by selecting events with a coincident triton cluster. While this removes uncertainty due to background subtraction, the cost is a reduction by a factor of 80  in the final number of events.  The $^{10}$N decay-energy distribution obtained from the $p$+$^9$C subevents gated on longitudinal $|\cos\theta_p|>0.5$ and transverse $|\cos\theta_p| < 0.5$ decays are shown in Figs.~\ref{fig:N10}(a) and \ref{fig:N10}(b), respectively. Both distributions show the presence of a wide peak at $E_{T}\approx$3~MeV, but more than one state is expected to be present in the region of $E_{T}$ sampled.

We have repeated the fit of these data based on the previous analysis where  two peaks with $R$-matrix lineshapes were assumed. The lower-energy ground-state peak, which is largely unresolved, was taken as an $s$-wave resonance and is expected to be vary wide \cite{Hooker:2017}.  The dominant  higher-energy peak was assumed to be  a $p$-wave resonance as it was also excited in proton pickup reactions with a fast $^9$C beam \cite{Charity:2021b}.
The original analysis did not include any background contribution from the prompt protons. However in the present fit, a background shape is determined from the same recipe as used in the other cases in this work. 
With  the fitted background now comprising 54\% of the yield, the relative strength of the  $s$-wave resonance is significantly reduced  and its parameters are not well constrained.

With the presence of a background component, which can now explain the high-energy tails of the experimental distributions, the main $p$-wave resonance is much narrower in the present fit. Its intrinsic lineshape was characterized by a maximum value of $E_{T}$=2.77(2)~MeV and a  FWHM of 2.45(7)~MeV in the original fit. In the present fit, the  
latter has shrunk to 1.57(12)~MeV and the maximum has shifted to $E_{T}$=2.81(3)~MeV.  This example again clearly shows the importance of understanding the background from prompt protons.  We have assumed isotropic decay of the $^{10}$N parent in the fits with equal yields for transverse and longitudinal decays (before the effects of the detector acceptance). This is consistent with distributions in Fig.~\ref{fig:Erel_10N}, but an anisotropy of magnitude similar to that in Fig.~\ref{fig:ang_pB8}(c) would be difficult to observe with the present statistics.  

\section{Levels in $^9$N}
For the $^{10,11}$N resonances, one could argue that it would be more efficient to study these via 1$n$ and 2$n$ knockout reactions from a $^{12}$N beam.  This method of production would negate the possibility of contributions from $^{11,12}$O decay. However, if $E$-$\Delta E$ detectors are still used, then the misidentification of carbon isotopes would still be a problem. In these neutron-knockout reactions, there are no clusters to tag on in order to the purify the spectra. Alternatively, spectrometers can be used to separate these isotopes with less misidentification issues.

One could consider producing the next lightest nitrogen isotope, $^9$N, via a 3$n$ knockout reaction, but the yield is expected to be quite small. In a companion letter \cite{Charity:2023}, the first evidence for the production of $^9$N state(s) is presented. These events were produced with  the $^{13}$O beam when a total of one proton and three neutrons were removed from the projectile. 
The nuclide $^9$N decays into the 5$p$+$\alpha$ exit channel and these decay products can be identified cleanly in the $E$-$\Delta E$ spectra, so there are no misidentification issues as in the heavier nitrogen isotopes. Peaks in the invariant-mass spectra were observed when a $^8$C(g.s.) intermediate state [$\Gamma$= 130(50)~keV] was identified in the 4$p$+$\alpha$ subevents.  In this section, we will examine to what extent contributions from decay of higher-$Z$ resonances contribute to the observed spectra and the background from prompt protons.

As we are considering an initial two-body $p$+$^8$C(g.s.) resonance, the issues here are very similar to those for the heavier nitrogen isotopes discussed in Secs.\ref{sec:N11} and \ref{sec:N10}. Contributions to the spectrum from higher-$Z$ resonances would be associated with the creation of the yet-to-be-observed $^{10}$O nuclide via a 3$n$ knockout reaction. We expect the ground state, and possibly some excited states, of $^{10}$O to 2$p$ decay to $^8$C(g.s.). If only one of these initial protons is detected, then such events would contribute to the $p$+$^8$C(g.s.) yield.  A total of eleven 6$p$+$\alpha$ events were detected in this experiment. With 15 4$p$+$\alpha$ subevents for each of these events, there are 165 values of $^8$C invariant mass. Only one of these lay in the invariant-mass gate used to select $^8$C(g.s.) intermediate states in Ref.~\cite{Charity:2023}. Given the profusion of these subevents, it is not clear if this  event corresponds to some background or is a real $^8$C(g.s.) intermediate. Even if the latter is true, we may still be observing the production of an $^{8}$C(g.s) resonance  in the first step of the reaction and not the decay of a  $^{10}$O resonance.

Assuming we have one real 2$p$+$^8$C(g.s.) event detected, then accounting for the proton detection efficiency, we should expect 2 events where one of these initial protons flies in between the HiRA detectors. With a total of 186 detected 5$p$+$\alpha$ events, then the contribution from the decay of $^{10}$O can be ignored. This also confirms our expectation about the small magnitude of the 3$n$ knockout cross section.

The observed $p$+$^8$C(g.s.) events are produced either from the decay of a $^9$N resonance or both the proton and $^8$C resonance are produced in the initial prompt fragmentation step.  The shape of the latter contribution to the invariant-mass spectrum can be estimated from our event-mixing recipe. A proton from one $p$+$^8$C(g.s.) event is mixed with the $^8$C(g.s.) decay products from a different event.  These events are weighted based on the $E_{p-^8\mathrm{C}}$ decay energy. While some $d$ and $t$ clusters were detected in coincidence with the 5$p$+$\alpha$ events, the numbers were not sufficient to construct useful correlation functions.  The weighting was therefore based on the correlation functions obtained for the heavier nitrogen isotopes in Secs.~\ref{sec:N11} and \ref{sec:N10} which are very similar in shape to each other.

\begin{figure}   
\centering
\includegraphics[scale=0.45]{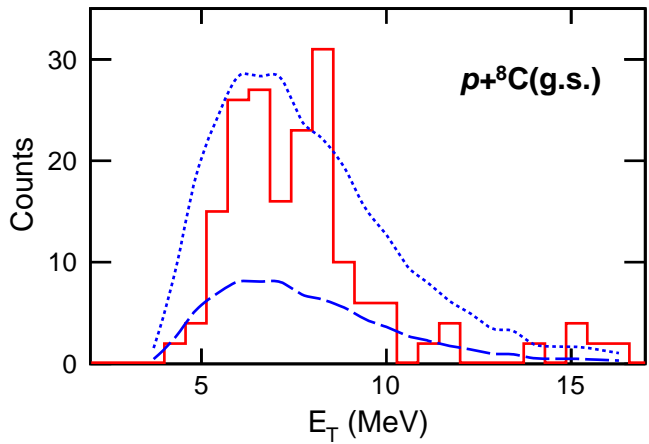}
%home/Boron8/Oxygen11/sortcode_addback/tree/N9/5pa/fit2peakBack/back.C 
\caption{\label{fig:N9} 
Histogram shows the experimental distribution of 5$p$+$\alpha$ total decay energy obtained from $p$+$^8$C(g.s.) events in  Ref.~\cite{Charity:2023}. The two blues curves show the background from promptly emitted protons as determined from the weighted mixed events. The upper dotted curve has been normalized to a similar maximum value as the experimental data to emphasis its difference in shape to the experimental result.
The lower dashed curve is normalised to reproduce just the high-energy and lowest-energy tail regions which appear to be dominated from this background.}
\end{figure}

Figure~\ref{fig:N9} compares the experimental 5$p$+$\alpha$ decay energy spectrum for $p$+$^8$C(g.s.) events to the predicted background from the prompt protons. The dotted curve shows this latter distribution normalized to give approximately the same maximum value as the experimental data. This curve emphasises that the shape of  this background distribution is too wide to explain the experimental data and thus we must have contributions from $^9$N resonance(s). The dashed curve gives a more reasonable normalization with this background adjusted to reproduce the yield in the high-energy ($E_T>$10~MeV) and low-energy ($E_{T}<$~MeV) tails.
This normalization suggests that the background accounts for roughly 40\% of the total yield which is similar to the results obtained in Fig~\ref{fig:N11}(b) and Fig.~\ref{fig:N10} for the two heavier nitrogen isotopes.
Fits to this experimental distribution with this background distribution and either one and two peaks from $^9$N resonances are presented in Ref.~\cite{Charity:2023}.

\section{Levels in $^{12}$O}
Finally we will look at $^{12}$O levels formed via single neutron knockout reactions and which decay by prompt 2$p$ emission. While neutron knockout reactions are clean in themselves, in the experimental invariant-mass spectrum of detected 2$p$+$^{10}$C events there will also be contributions from reactions where the $^{10}$C fragment is produced promptly in the first step and from the decay of $^{11}$N resonances (Sec.~\ref{sec:N11}). 

\begin{figure}   
\centering
\includegraphics[scale=0.43]{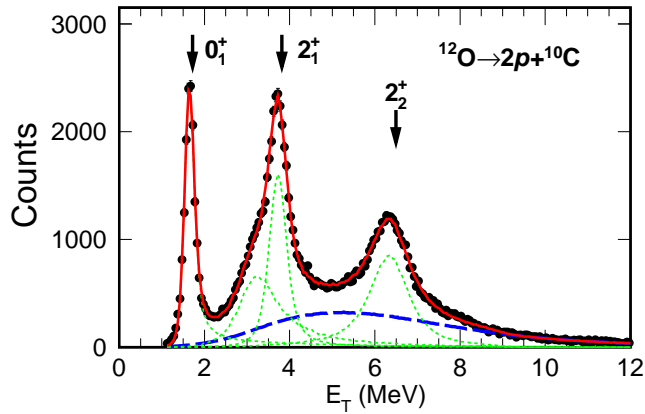}
%home/Boron8/Oxygen11/sortcode_addback/tree/O12/fitBackdTransNarrow/chi.C and fit.root 
\caption{\label{fig:O12} 
Data points show the distribution of $^{12}$O decay energy obtained from the detected 2$p$+$^{10}$C events where the emission angle $\theta_{10}$ of the $^{10}$C fragment in the parent's reference frame is constrained to $|\cos\theta_{10}|<$0.2. The solid red curve is a fit to this distribution with four peaks (dotted green curves) and the prompt-proton background (dashed blue curve). Arrows indicate the energies of levels with known spin/parity.}
\end{figure}

The decay-energy spectrum obtained from the 2$p$+$^{10}$C events is shown in Fig.~\ref{fig:O12}. As there are some relatively narrow levels present, and there is adequate statistics, we have gated on events where the $^{10}$C fragment is emitted transversely ($|\cos\theta_{10}|<$0.2) to improve the experimental resolution \cite{Charity:2019a}. The angle $\theta_{10}$ is the emission angle of the $^{10}$C residual in the $^{12}$O reference frame. Three strong peaks are present which are labeled by their spin/parity in the figure. The middle peak also has a low-energy shoulder associated with another resonant decay. 

This spectrum was originally fit in Ref.~\cite{Webb:2019a} with a background component which was parameterized by a simple function with the only requirement being that its energy dependence is smooth. In this work, the background is generated from our weighted event-mixing procedure.  Initially this background was constructed from detected $p$+$^{10}$C events where we mixed in another proton from another such event. The weighting was based the relative energy between this new proton and the $^{10}$C fragment using the same weighting as for the background in the $p$+$^{10}$C spectrum in Fig.~\ref{fig:N11}.
However, these $p$+$^{10}$C events contain significant contributions from $^{12}$O states formed in 1$n$ knockout reaction where there are no prompt protons. So mixing in a prompt proton to this contribution may be problematic, and hence it seemed more reasonable to mix the $p$+$^{10}$C subevents in the detected $d$+$p$+$^{10}$C events where it was argued that this component was missing (Sec.~\ref{sec:N11}).  The differences in the shapes of the two generated background distributions are small, and fits of the experimental spectrum  with both of these  backgrounds have been made in order to gauge the  uncertainties in fitted parameters. 

The solid red curve in Fig.~\ref{fig:O12} shows the best 4-peak fit with the second background prescription. The background (dashed blue curve) represents 30\% of the detected yield increasing to 34\% if the gate on $\theta_{10}$ is relaxed. This fraction is of similar order to those from the other channels considered, indicating that this is not  a particularly ``clean'' channel in this respect.

Another difference to the original fit in Ref.~\cite{Webb:2019a} is that instead of Breit-Wigner line shapes we have now used $R$-matrix line shapes for a simple di-proton emission scenario as used in fitting the neighboring $^{11}$O resonances \cite{Webb:2019}.  This difference was most important for the 0$^+_1$ ground state  which now has little background  under its peak and  the asymmetric lineshape from this $R$-matrix formalism allows for a much better fit. A comparison of the fitted intrinsic widths and peak decay energies is listed in Table~\ref{tbl:O12}. The fitted widths are now all larger with the new background prescription. Some of the small shifts in energy of the states between the two analyses is due to refinements to the energy and angular calibrations. The 0$^+_1$ ground-state energy is now consistent with the value of $E_{T}$=1.638(24)~MeV from the earlier invariant-mass study of Ref.~\cite{Jager:2012}. 

The low-energy shoulder next to the more prominent 2$^+_1$ peak could be due to yield from the decay of the 0$^+_2$ state which is expected in this vicinity \cite{Suzuki:2016} or a decay branch of the 2$^+_2$ state that proton decays to the 2$^+$ excited state of $^{10}$C which then gamma decays to the ground state. If the latter scenario is correct, then the intrinsic width of this fitted shoulder peak should be consistent with that extracted for the 2$^+_2$ state  and the difference in their mean energies should be 3.353~MeV, the energy of the emitted $\gamma$ ray.  Both are correct and the latter is true to within 191(127)~keV making this scenario the most reasonable of the two.

\begin{figure}   
\centering
\includegraphics[scale=0.45]{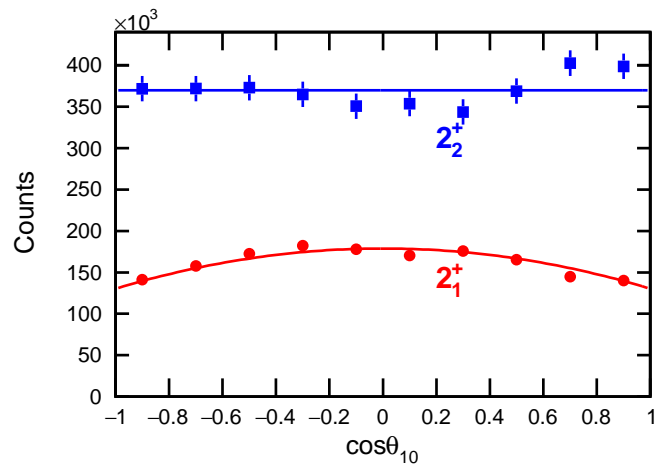}
%home/Boron8/Oxygen11/sortcode_addback/tree/O12/fitBack_p/ang.C 
\caption{\label{fig:ang_12O} 
Distribution of the $^{10}$C decay angle in the parent $^{12}$O fragment's reference frame for the two 2$^+$ states in $^{12}$O.}
\end{figure}

With the large statistics for this channel, it is possible to look at the decay angular distributions of the 2$^+$ states from fitting different $\theta_{10}$ gates. The resulting angular distributions are shown in Fig.~\ref{fig:ang_12O} and, like the result for the $^9$C$\rightarrow p$+$^8$B decay in Sec.~\ref{sec:C9}, the results have the expected symmetry about $\cos\theta_{10}$=0. 
The result for the 2$^+_2$ state is consistent with a flat distribution, i.e. isotropic emission. However the 2$^+_1$ state has a small anisotropy.  Significant spin alignment in nucleon  knockout reactions have been predicted when there is a bias on the detected momentum distributions of the products \cite{Hansen:2003}. In this experimental study, large transverse momenta for the $^{12}$O resonances are needed in order to get the $^{10}$C fragments out to the inner angles subtended by the HiRA array. As yet there is no theoretical treatment of the angular distributions of the residual fragment for spin-aligned prompt 2$p$ emitters.

\begin{table}
\caption{\label{tbl:O12} Comparison of level parameters for $^{12}$O states obtained in the present analysis to those from the original analysis in Ref.~\cite{Webb:2019a}. The uncertainty in the parameters in the present fit accounts for differences due to the two different background prescriptions described in the text. In addition, a systematic uncertainty in $E^*$ of about 10 keV is expected from past experience \cite{Charity:2019}.
For asymmetric line shapes used in the present fit, there can be differing definitions for the centroid and width. Here we have chosen $E^*$ as the value corresponding to the peak of the intrinsic line shape and the width is represented by its FWHM.}
\centering
\begin{ruledtabular}

\renewcommand{\arraystretch}{1.5}
\begin{tabular}{ccccc}
         &           \multicolumn{2}{c}{present analysis} & \multicolumn{2}{c}{original analysis} \\
$J^\pi$ &  $E^*$  & FWHM & $E^*$ & FWHM\\
         &   (MeV) & (keV) & (MeV) & (keV) \\
\hline
0$^+_1$ & 1.636(2) & 107(13) & 1.718(15) & 51(19)\\
2$^+_1$ & 3.733(38) & 262(34) & 3.817(18) & 155(15)\\
2$^+_2$ & 6.395(21) & 900(273) & 6.493(17) & 754(25) \\
shoulder & 3.23(12) & 1019(426) & 3.519(67) & 980(182) \\

\end{tabular}
\end{ruledtabular}
\end{table}

\section{Discussion and Conclusions}
It is clear from the channels we examined that the projectile fragmentation process has an initial prompt step creating both bound nuclei and resonances.  Similar results were obtained in the fragmentation of a $^{20}$Ne projectile in Ref.~\cite{Charity:1995}.  With so many resonances produced in the fragmentation process, the application of the invariant-mass method can be quite fruitful even for just a single projectile species. We have shown that the very small angles probed by the HiRA array ($<$12$^\circ$)  concentrate the decay products from low-lying resonances. Promptly emitted protons produced in the first step are typically emitted  to much larger angles. This is consistent with studies of one and two-proton knockout.  In  single-proton knockout reactions from  $^9$C and  $^8$B projectiles at the even higher energies of 98~MeV/$A$ and 87~MeV/$A$, respectively,  where the prompt protons should be more forward focused, significant yields of the knockout protons were found out to angles of 40$^\circ$ and beyond \cite{Bazin:2009}.  Significant yields at similarly large angles have also been observed in the 2$p$ knockout from a 93-MeV/$A$ $^{28}$Mg beam \cite{Wimmer:2012}.

In searching for proton resonances in invariant-mass spectra, even through most of the prompt protons are found at larger angles, there is always some significant background from coincidences with the remaining prompt protons at small angles. We have shown that understanding the shape of this component in the invariant-mass spectrum is very important in fitting the observed resonance peaks. In particular, incorrect background shapes can lead to significant errors in extracting the widths of  wide resonances. 

While event mixing, which assumes no correlations between the mixed particles, has been considered in constructing backgrounds in the past, this does not work at the smaller decay energies. Based on observations with light clusters which are predominately produced in the first step of the fragmentation process, we observe a suppression of such particles in the phase space near the cluster. The correlation functions obtained for $d$ and $t$ clusters, as a function of their energy relative to the resonance, can be used to estimate the suppression for prompt protons. Our recipe for estimating the $E_{T}$ dependence of the background from prompt protons involves mixing events, but weighting these mixed events by the correlation function obtained from coincident $d$ and $t$ clusters.   The magnitude of this background is obtained from fits to the invariant-mass spectrum. In many cases, it is largely determined from the magnitude of the high-energy tail and sometimes the low-energy tail of the experimental spectrum when these are free of resonances.  This background comprised 30-54\% of observed yield and was not strongly dependent of how many protons were removed from the projectile to create the resonance. While there are clean reactions such a 1$n$, 2$n$ knockout and inelastic excitation that produce no prompt protons, these cannot be isolated by themselves in invariant-mass spectroscopy. All exit channels in projectile fragmentation contain contributions from the  non-resonant protons.

For some channels there was additional background from multi-proton decays of resonances in higher-$Z$ nuclei.  We have presented two ways to dealing with this. One involves using detected events with extra protons, which have greater relative contributions from these higher-$Z$ resonances, and constructing the invariant-mass spectra of their sub-events to predict this background component. An alternative procedure was also presented where one detects proton resonances in coincidence with a prompt light cluster. If the presence of a cluster precludes the formation of the higher-$Z$ resonances in the first step of the reaction, then the proton-plus-residual subevent is free of this background. However the price paid to cleanly remove this background is significantly reduced statistics. This price was even larger in this study due to the loss of identification of $d$ and $t$ clusters that punch through the $\approx$4~cm CsI(Tl) $E$  detectors. Thicker detectors such as the upgraded HiRA10 array with 10~cm-long CsI(Tl) crystals \cite{Sweany:2021} would help alleviate this deficiency.

In conclusion, projectile fragmentation reactions ranging from single-nucleon removal to multi-fragment creation produce a large variety of resonances which can be studied with the invariant-mass method. However, proper treatment of backgrounds from non-resonant particles is required to accurately constraint resonance parameters. A event-mixing scheme where  mixed events are suppressed at small invariant-masses  was found to give a good description of this background. These fragmentation reactions can be used to study the furthest limits of the chart of nuclides well beyond the drip lines.

\begin{acknowledgments}
This material is based upon work supported by the U.S.
Department of Energy, Office of Science, Office of Nuclear
Physics under Award No. DE-FG02-87ER-40316.

\end{acknowledgments}

%\bibliography{references}% Produces the bibliography via BibTeX.


\end{document}